\documentclass[conference]{IEEEtran}
\IEEEoverridecommandlockouts
\usepackage{cite}
\usepackage{amsmath,amssymb,amsfonts}
\usepackage{algorithmic}
\usepackage{graphicx}
\usepackage{textcomp}
\usepackage{xcolor}

\newcommand{\tacotron}{\text{\emph{Tacotron} }}
\newcommand{\wavenet}{\text{\emph{WaveNet} }}

\newcommand{\tts}{\text{\emph{TTS} }}

\def\BibTeX{{\rm B\kern-.05em{\sc i\kern-.025em b}\kern-.08em
   T\kern-.1667em\lower.7ex\hbox{E}\kern-.125emX}}

\DeclareGraphicsExtensions{.eps,.png,.jpg}
\graphicspath{ {./Figures/} }
\begin{document}

\title{Taco-VC: A Single Speaker Tacotron based Voice Conversion with Limited Data}

\author{\IEEEauthorblockN{Roee Levy-Leshem}
\IEEEauthorblockA{\textit{School of Electrical Engineering} \\
\textit{Tel Aviv University}\\
Tel Aviv, Israel \\
roeelev1@mail.tau.ac.il}
\and
\IEEEauthorblockN{Raja Giryes}
\IEEEauthorblockA{\textit{School of Electrical Engineering} \\
\textit{Tel Aviv University}\\
Tel Aviv, Israel \\
raja@tauex.tau.ac.il}
}

\maketitle
\begin{abstract}
This paper introduces \emph{Taco-VC}, a novel architecture for voice conversion based on \emph{Tacotron} synthesizer, which is a sequence-to-sequence with attention model. The training of multi-speaker voice conversion systems requires a large number of resources, both in training and corpus size. \emph{Taco-VC} is implemented using a single speaker \emph{Tacotron} synthesizer based on Phonetic PosteriorGrams (\emph{PPG}s) and a single speaker \emph{WaveNet} vocoder conditioned on mel spectrograms. 
 To enhance the converted speech quality, and to overcome over-smoothing, the outputs of \emph{Tacotron} are passed through a novel speech-enhancement network, which is composed of a combination of the phoneme recognition and \emph{Tacotron} networks. 
 Our system is trained just with a single speaker corpus and adapts to new speakers using only a few minutes of training data.
  Using mid-size public datasets, our method outperforms the baseline in the \emph{VCC} 2018 SPOKE non-parallel voice conversion task and achieves competitive results compared to multi-speaker networks trained on large private datasets.
\end{abstract}


\section{Introduction}
The purpose of voice conversion (\emph{VC}) is to convert the speech of a source speaker into a given desired target speaker. A successful conversion preserves the linguistic and phonetic characteristics of the source utterance while keeping naturalness and similarity to the target speaker. 
\emph{VC} can be applied to various applications, such as personalized generated voice in text-to-speech (\emph{TTS}) \cite{Latorre2014}, speaking aid for people with vocal impairments \cite{Erro2015} and speaker verification spoofing \cite{Wu2015}.

A wide range of approaches exists for the \emph{VC} task. Some use a statistical parametric model such as Gaussian mixture models (\emph{GMM}) to capture the acoustic features of the source speaker and create a conversion function that maps to the target speaker \cite{Stylianou1998,Toda2007}. Recently, several deep learning based solutions have been provided and successfully led to a better spectral conversion compared to the traditional GMM-based methods. Various network architectures are employed such as feed-forward deep neural networks \cite{Chen2014,Saito2017} recurrent neural networks (\emph{RNN}) \cite{Ramos2017,Sun2015}, generative adversarial networks (\emph{GAN}) \cite{Saito2018b,Kaneko2018}, and variational autoencoder (\emph{VAE})\cite{Saito2018a, Kameoka2019}.

\begin{figure}[htbp]
\centerline{\includegraphics[width=0.45\textwidth]{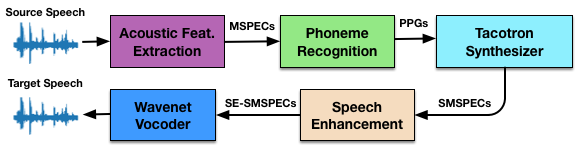}}
\caption{\emph{Taco-VC} Conversion Process.}
\label{fig:eval_vc}
\end{figure}

The converted speech of a \emph{VC} system is measured by three main quality parameters: (1) Prosody preservation of the source speech, (2) naturalness, and (3) target similarity. Recent research demonstrates successful prosody preservation when using \emph{VC} based phonetic PosteriorGrams (\emph{PPG}s) \cite{Sisman2018}. \emph{PPG}s represent the posterior probability of each phonetic class per single frame of speech. The \emph{PPG}s are obtained from speaker-independent automatic speech recognition (\emph{SI-ASR}) network, therefore considered as \emph{SI} features \cite{Sun2016a}. The quality of the converted speech is profoundly affected by the vocoder used in the speech synthesis system. Recently, \wavenet vocoder \cite{Oord2016} became highly popular and is broadly used in \emph{VC}, providing high quality converted waveforms \cite{Kobayashi2017, Liu2018a}.

\emph{TTS} research has gained significant progress over the last years, mainly due to the adaptation of sequence-to-sequence (\emph{Seq2Seq}) models such as \emph{Tacotron} \cite{Wang2017a, Shen2018}. \emph{Seq2Seq} methods are also used for \emph{VC}, among them, the multi-speaker \emph{SCENET} model \cite{Zhang2019} contains an encoder-decoder with attention, which predicts target \emph{MSPEC}s from source \emph{MSPEC}s and bottleneck features. The \emph{Parrotron} \cite{Biadsy2019} and the work from \cite{Zhang2019a} also describe the usage of \emph{Tacotron} for \emph{VC} purposes, however, they do not use prosody preserved features and require text or phonemes during training.

In this work, we propose \emph{Taco-VC}, a four stages architecture for high quality, non-parallel, many-to-one \emph{VC}. Its main advantage is that it requires a corpus of only a single speaker for training, and can easily be adapted to other speakers with limited training data. Inspired by the recent success of \emph{TTS} models, we base our \emph{VC} system on the architecture of \emph{Tacotron} \cite{Wang2017a}, which provides high quality and natural speech using a \emph{Seq2Seq} synthesizer with attention mechanism \cite{Bahdanau2014}, and \emph{WavenNet} vocoder. As can be seen in Fig. \ref{fig:eval_vc}, Phonetic PosteriorGrams (\emph{PPG}) are extracted from a phoneme recognition (\emph{PR}) model to preserve the prosody of the source speech. Using a single speaker \emph{Tacotron} synthesizer, we synthesize the target mel-spectrograms (\emph{MSPEC}) directly from the \emph{PPG}s. The synthesized \emph{MSPEC}s (\emph{SMSPEC}) pass through a speech enhancement network (\emph{Taco-SE}), which outputs the speech enhanced \emph{SMSPEC}s (\emph{SE-SMSPEC}). Finally, a single speaker \emph{WaveNet} vocoder generates the predicted audio from the \emph{SE-SMPSEC}s. We use the same acoustic features (80-band \emph{MSPEC}s) in our different networks as it leads to a high-quality conversion in terms of similarity to the target speaker \cite{Chen2018}. It also allows to train the different networks independently and combine them to generate the final target audio.

The main contributions of this work are: (1) a scheme that relies on a single-speaker \emph{Tacotron} and \emph{WaveNet}, and adapts successfully to other target speakers with limited training data; (2) a novel approach for speech enhancement, which handles over-smoothing and noise using a joint training of the \emph{PR} and \emph{Tacotron} synthesizer without over-parameterization of the model due to weight sharing; (3) a \emph{VC} architecture that uses only public and mid-size data, and outperforms the existing baselines. It also shows competitive results compared to other multi-speaker \emph{VC} networks trained on private and much larger datasets. To the best of our knowledge, \emph{Taco-VC} is the first \emph{VC} system that presents a successful adaptation of single speaker networks to other speakers with limited data.

The paper is organized as follows: Section \ref{VoiceConversionNetwork} describes  \emph{Taco-VC} model and its adaptation process to new speakers. Section \ref{Experiments} reports the experiments and results, showing the advantages of the proposed approach. Section \ref{Conclusion} concludes the paper. 
\section{The Voice Conversion Network}\label{VoiceConversionNetwork}
Fig. \ref{fig:train_vc} presents the four components of our \emph{VC} system. We provide next details on each of them. 


\subsection{The Phoneme Recognition Network}

We use Phonetic PosteriorGrams as our prosodic preserving features. The \emph{PPG}s are extracted using a \emph{PR} network. This network architecture choice is made with two main goals: (1) Provide the ability to extract \emph{PPG}s at the frame level; (2) Allow joint training with the speech synthesis network. We use a convolutional neural network (\emph{CNN}) based \emph{PR}, which is easy to train as it suffers less from vanishing gradients issues \cite{Bengio1994} during training compared to \emph{RNN} and it can be integrated with \emph{Tacotron} synthesizer as part of the encoder. The \emph{PPG}s are taken from the last fully connected layer before the \emph{CTC} loss \cite{Graves2014}, which is employed in our network training.
	
\begin{figure}[htbp]
\centerline{\includegraphics[width=0.5\textwidth]{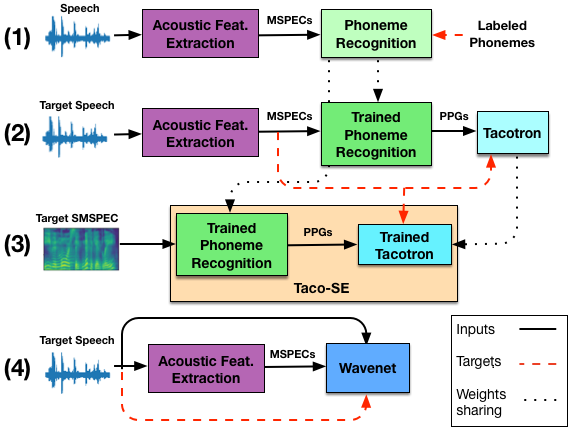}}
\caption{The training of our model consists of four steps: (1) Phoneme recognition training, 
(2) \emph{Tacotron} Synthesizer training, (3) Speech enhancement (\emph{Taco-SE}) training, (4) \emph{WaveNet} training.}
\label{fig:train_vc}
\end{figure}	
	
Fig. \ref{fig:train_vc}(1) shows the \emph{Seq2Seq} training process of the \emph{PR} network with the \emph{MSPEC}s as the inputs and the phoneme labels as the targets. This network has the same structure of \cite{Zhang2016} except of the following changes: (1) We use the \emph{Leaky-ReLU} non-linearity \cite{Maas2013} instead of \emph{Maxout} to reduce the number of parameters; (2) We add batch-normalization \cite{Ioffe2015} after each non-linear activation in the convolution layers to increase network stability; (3) For compatibility with  \emph{Tacotron} and \emph{WaveNet} networks, the raw audio input of the \emph{PR} network is transformed into \emph{MSPEC}s instead of mel-cepstral coefficients.

The performance of our \emph{PR} network is measured by phoneme error rate (\emph{PER}). It achieves 17.5\% \emph{PER} on the core test set, which improves over the 18.2\% of the network in \cite{Zhang2016}.


\subsection{The Speech Synthesis Network (\emph{Tacotron})}

Inspired by the success of \emph{Tacotron} in the fields of \emph{TTS}, we propose a single speaker \emph{Tacotron} \emph{Seq2Seq} model with attention mechanism to predict \emph{MSPEC}s directly from the \emph{PPG}s extracted by the \emph{PR} network for the entire target speech corpus. While \tts systems are trained with pairs of $<Text, Audio>$, for \emph{VC} purposes, \tacotron  is trained with $<PPG, Audio>$ pairs. Fig. \ref{fig:train_vc}(2) shows the \emph{Seq2Seq} training of \emph{Tacotron}. The \emph{PPG}s are the single input of the network, while the \emph{MSPEC}s and linear-spectrograms are used as the target.

Our synthesis network has the same structure and loss function as the original \emph{Tacotron} \cite{Wang2017a} except of the following changes: (1) The Pre-net of the encoder \emph{CBHG} is fed directly with \emph{PPG}s instead of text; (2) While the original \tacotron uses teacher forcing mode in the training process, we use linearly decayed scheduled sampling \cite{Bengio2015} with a final sampling rate of 0.33 for true samples, which helps to increase the quality of the generated \emph{MSPEC}s, especially when adapting the single speaker model to a limited-size train set; (3) As the source utterance length is known, it can be used as the "stop-token" of the decoder, using the fact that the target utterance has the same length as the source utterance. We have found that constant stop-token helps to get more stable outputs in the generation process.


\subsection{The Speech Enhancement Network (\emph{Taco-SE})}

The \emph{PR} network and \tacotron are trained separately on different corpora. We have found that the synthesized \emph{MSPEC}s tends to be over-smoothed in the mid-high harmonics. Moreover, the over-smoothing artefacts get worse when adapting \emph{Tacotron}, which is trained on a single speaker speech corpus, to a different speaker with a limited train set.

To address these artefacts, we add another network, \emph{Taco-SE}, which is a concatenated network comprising of the trained \emph{PR} ($P(\bullet)$) connected to the trained \tacotron ($T(\bullet)$), without over-parameterization of the model due to weights sharing (see Fig. \ref{fig:train_vc}(3)). After initialization, \emph{Taco-SE} is trained using only the \tacotron loss $L_T$. As the purpose of \emph{Taco-SE} is to enhance the quality of the \emph{SMSPEC}s, we generate for the entire corpus, using the first two networks, the \emph{SMSPEC} of each utterance. To train the network to increase the quality, we require it to generate the true \emph{MSPEC}, denoted as $y$, from the \emph{SMSPEC}s, denoted as $\hat{y}$. We also require it to provide this output if $y$ is given as we want the \emph{Taco-SE} to preserve high-quality inputs.

To summarize, \emph{Taco-SE} is trained on the pairs $<y,y>$ and $<\hat{y},y>$, each with probability 0.5. The first corresponds to retaining the quality by recovering the true target signal given as an input, and the second aims at estimating the target speech signal from a synthesized one with the goal of improving the quality of the network. This leads to the following loss:
\begin{equation}
L_{Taco-SE}=L_T (T(P(y)),y)+L_T (T(P(\hat{y})),y).
\end{equation} 

\begin{figure}[htbp]
\centerline{\includegraphics[width=0.5\textwidth]{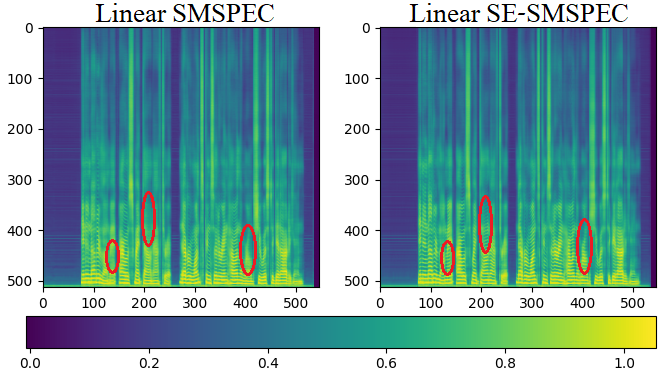}}
\caption{Linear \emph{SE-SMSPEC} and \emph{SMSPEC} Comparison.}
\label{fig:linspec}
\end{figure}

As can be seen in Fig. \ref{fig:linspec}, the primary enhancement of \emph{Taco-SE} is being reflected in the mid-higher harmonics (marked by red circles), while in the lower harmonics, there are merely no changes. The \emph{SE-SMSPEC} contains much better-resolved harmonics compare to the \emph{SMSPEC}.


\subsection{The Vocoder Network (\emph{WaveNet})}

The conditional \wavenet vocoder aims at reconstructing the target raw waveforms from \emph{MSPEC}s. For conditioning the \emph{MSPEC}s, we add local conditioning to the gated units. Since the \emph{MSPEC} is sampled with a lower sampling frequency compares to the raw waveform, we add learnable up-sampling convolutional layers that map it to a new time series with the same resolution of the raw waveform.

We use the implementation and parameters of \wavenet from \cite{Yamamoto}. As Fig. \ref{fig:train_vc}(4) shows, for the \wavenet training, we use the same single speaker speech corpus used for both \tacotron and \emph{Taco-SE}. Also, for local conditioning of \emph{WaveNet}, we use the same \emph{MSPEC}s features that are used for the rest of the networks.

\subsection{System Adaptation}

Another aspect of speech synthesis systems in general and \emph{VC} systems, in particular, is the ability to adapt to new speakers given limited training data. \emph{TTS} models are usually trained on large datasets with multiple-speaker support. There are two main strategies for adapting to other target speakers: (1) Using a speaker embedding in multi-speaker systems \cite{Nachmani2018}; and (2) model adjustment by fine-tuning of a multi-speaker \emph{SI} model to a target speaker, which leads to  better results in terms of target similarity \cite{Arik2018}. Such multi-speaker networks require longer training phases, complex networks with a large number of parameters, and much larger training sets. The usage of model adjustment was also explored for \emph{VC} systems, such as \cite{Sisman2018,Liu2018a} that train a multi-speaker \emph{SI} \wavenet vocoder and adapt it to new target speakers.

We use the model adjustment by the fine-tuning technique for the adaptation process, as done in \cite{Liu2018a}, and explore how a single speaker system will adapt to other speakers with limited data. The trained \tacotron is fine-tuned on the new target's training data with linearly decayed scheduled sampling. \emph{Taco-SE} is fine-tuned in the same way as \tacotron and uses \emph{SMPSEC}s that are generated for every utterance in the new target training set by the fine-tuned \emph{Tacotron}. \wavenet is also fine-tuned on the new target training set. Since the \emph{PR} network is speaker independent, it does not require an adaptation. 
\section{Experiments}\label{Experiments}

\subsection{Experimental Setups}

The \emph{PR} model is trained using the \emph{TIMIT} corpus \cite{Garofolo1993}. All the 462 speakers training set is used except the SA recordings. The sampling rate of the \emph{TIMIT} is 16 kHz with a 16-bit resolution. For having alignment with the rest of the networks, we up-sampled it to 22050 Hz. \emph{Tacotron}, \emph{Taco-SE}, and \wavenet are trained using the public LJ Speech corpus \cite{KeithIto2017}, which consists of 13,100 utterances from a single female speaker. The total length of the corpus is approximately 24 hours. All of the utterances are recorded with a sampling rate of 22050 Hz and a 16-bit resolution. The acoustic features used for all of the systems are 80-band \emph{MSPEC}s extracted using Hann windowing of 1024-samples Short Time Fourier Transform, and 256-samples step size. The mel filter-bank base is computed in the range of 125 to 7600 Hz. 

We evaluate our system on the VCC2018 SPOKE task \cite{Lorenzo-Trueba2018}, which is a non-parallel \emph{VC} task. It has an English speech dataset, containing two males (VCC2TM1, VCC2TM2) and two females (VCC2TF1, VCC2TF2) target speakers and two males (VCC2SM3, VCC2SM4) and two females (VCC2SF3, VCC2SF4) source speakers. Each speaker has the same 81 content utterances for training, and 35 utterances for testing.  The whole training set is approximately 5 minutes of speech per target speaker. All of the utterances are recorded with a sampling rate of 22050 Hz and a 16-bit resolution.

\begin{figure}[htbp]
\centerline{\includegraphics[width=0.45\textwidth]{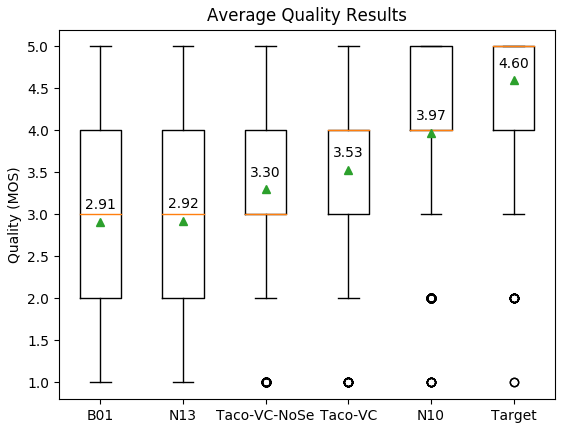}}
\caption{Total Average Quality (Naturalness) MOS of the five evaluated networks and target speech. The triangle value is the mean. The bold line is the median.}
\label{fig:chart:nat_all}
\end{figure}

\emph{Tacotron} and \emph{Taco-SE} are trained with a batch size of 5 and optimized using Adam optimizer with a linearly decayed learning rate with an initial value of 0.002 for \emph{Tacotron} and 0.0005 for \emph{Taco-SE}. We use reduction factor $r=3$ as it leads to the best attention alignment. To adapt the different networks, we fine-tuned the trained \emph{Tacotron} and \emph{Taco-SE} for each of the target speakers for another 10,000 steps, using linearly decayed scheduled sampling as for the initial training. \wavenet is fine-tuned with another 20,000 steps.

For subjective evaluation, we use the mean opinion score (\emph{MOS}) of naturalness and target similarity. Both evaluations are conducted using the Amazon Mechanical Turk framework. We compare our test utterances to the published, submitted test utterances of the VCC2018. We also do an ablation study by removing the \emph{Taco-SE} network. The tested models are\footnote{Audio samples - https://roee058.github.io/Taco-VC/}:

\begin{itemize}
	\item B01 - The baseline system of VCC2018 is a vocoder-free system based on a  \emph{GMM} conversion model \cite{Kobayashi2018}.
	\item N10 - The best system in both the similarity and naturalness scores of VCC2018 \cite{Liu2018a}. It uses a \emph{DBLSTM} conversion model that converts \emph{STRAIGHT} extracted spectral features and $F_0$. The vocoder is a speaker-dependent multi-speaker \emph{WaveNet}. The networks are trained using iFlytek large private datasets. As we do not have access to this large corpus, this work has an inherit advantage.
	\item N17 - The second-best system in the similarity score of VCC2018 \cite{Wu2018}. The conversion model is \emph{DNN} based encoder-decoder trained on a parallel training set generated by \emph{TTS} from the non-parallel corpus. The vocoder is a speaker-dependent multi-speaker \emph{WaveNet}.
	\item N13 - The second-best system in the naturalness score of the VCC2018.
	\item \emph{Taco-VC} - Our proposed method, including \emph{Taco-SE} network.
	\item \emph{Taco-VC-NoSe} - Our proposed method without \emph{Taco-SE} network.
\end{itemize}

\begin{figure}[htbp]
\centerline{\includegraphics[width=0.45\textwidth]{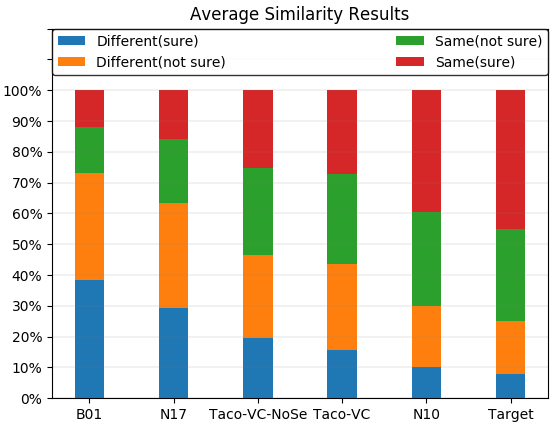}}
\caption{MOS of target similarity of the evaluated five networks and target speech.}
\label{fig:chart:sim_all}
\end{figure}


\subsection{Naturalness Evaluation}

In the naturalness evaluation, human subjects rate the quality of the different converted utterances. In each assignment, subjects rate six different utterances with the same content speech - N10, N13, B01, \emph{Taco-VC}, \emph{Taco-VC-NoSe}, and the original target. The quality rate is on a scale of 1 (Bad - Completely unnatural speech) to 5 (Excellent - Completely natural speech). The number of evaluation utterances is ten conversions per source with a total of 40 per target, and a total of 160 utterances per system. Every utterance gets ten votes. The utterances are presented in random order. Total of 128 different evaluators participated in the experiment with an average of 74 utterances ranks.

Fig. \ref{fig:chart:nat_all} shows the average \emph{MOS} for naturalness averaged on all pairs. The results indicate a significant effect of the \emph{Taco-SE} on the quality scores. The quality \emph{MOS} results indicate that in terms of subjective quality evaluation, \emph{Taco-VC} outperforms the baseline and gets the same median as N10, though using only a single speaker baseline. The quality gap between \emph{Taco-VC} and N10 can also be explained by the relatively high \emph{PER} of the \emph{PR} network.

\subsection{Target Similarity Evaluation}

In the target similarity evaluation, subjects rate the similarity of the different converted utterances to the target speaker utterances. The reference target utterance is chosen by random selection from the training set. In each assignment, subjects rate six different test utterances with the same content speech - N10, N17, B01, \emph{Taco-VC}, \emph{Taco-VC-NoSe}, and the original target. The similarity rate is on a scale of 1 (Different - absolutely sure), to 4 (Same - absolutely sure). We use the same utterances as in the naturalness evaluation. Total of 165 different evaluators participated in the experiment with an average of 57 utterances ranks.

Figure \ref{fig:chart:sim_all} shows the \emph{MOS} distribution for target similarity averaged on all pairs. For \emph{Taco-VC}, almost 60\% are ranked as similar to the target, while the baseline (B01) has less than 30\%. The real target utterances get the rank of 75\%. Note that the impact of \emph{Taco-SE} on the similarity score is minor compared to the naturalness case.
\section{Conclusion}\label{Conclusion}

This work presents \emph{Taco-VC}, a \emph{VC} system comprised of \emph{PR} network, \tacotron synthesizer, and \wavenet vocoder. It has the advantage that it can produce a high-quality speech conversion by just being trained on a single speaker large corpus and then be adapted to new speakers only using a small amount of data. We also introduce the speech enhancement network \emph{Taco-SE}, which might be of interest by itself, and describe how to enhance the synthesized mel-spectrograms only using the trained networks. We show in the \emph{MOS} experiments that our architecture, using public, single speaker training set, can adapt to other targets with limited training sets, and provide competitive results compared to multi-speaker \emph{VC} systems trained on private and much larger datasets.
We believe that the high error rate of the \emph{PR} network has a significant impact on the converted speech. As future work, we suggest adding more acoustic features to the generated \emph{PPG}s, or extract \emph{PPG}s from other speech recognition networks with lower error rates. Another possible future research direction is applying the \emph{Taco-SE} architecture (with a corresponding \wavenet for denoising \cite{Rethage2018}) to speech denoising tasks.

\bibliographystyle{ieeetr}
\bibliography{references}
\vspace{12pt}

\end{document}